\documentclass[graybox]{svmult}
\usepackage{mathptmx}
\usepackage{helvet}
\usepackage{courier}
\usepackage{type1cm}
\usepackage{makeidx}
\usepackage{graphicx}
\usepackage{multicol}
\usepackage[bottom]{footmisc}
\usepackage{amsmath,amssymb}
\usepackage{url}
\usepackage{booktabs}
\usepackage{array}
\usepackage{float}
\usepackage{placeins}
\makeindex

\begin{document}

\title*{Envelope Model for GTEx Data}
\titlerunning{Envelope Model for GTEx Data}
\author{Lisa Chang, Yijun Wang, and Shanshan Ding}
\authorrunning{L. Chang et al.}
\institute{Lisa Chang \at Department of Applied Economics and Statistics, University of Delaware, Newark, Delaware, \email{clisa@udel.edu}
  \and Yijun Wang \at Department of Applied Economics and Statistics, University of Delaware, Newark, Delaware,  \email{yjwang@udel.edu}
  \and Shanshan Ding \at Department of Applied Economics and Statistics, University of Delaware, Newark, Delaware,  \email{sding@udel.edu}}

\maketitle

\abstract{This article studies the envelope model for enhanced gene expression analysis using GTEx data. The study compares this advanced statistical approach with standard multivariate linear regression, demonstrating improved capacity for identifying significant genes and enhanced estimation efficiency. By partitioning response variation into material and immaterial components, the envelope model reduces estimation noise, leading to stronger statistical signals and smaller standard errors. These findings, as demonstrated in our tissue-specific and gene-group analyses, show that the envelope model can pick up on important biological signals that traditional methods tend to miss. This makes it a useful tool for genomic studies and GTEx data analysis.}

\section{Introduction}\label{sec:introduction}
The GTEx project, launched in 2010, analyzed up to 54 non-diseased tissue types from nearly 1,000 donors, using whole-genome sequencing to assess genetic variation and bulk RNA sequencing to measure gene expression and genetic variation (GTEx Consortium, 2020). Through this effort, the project identified expression quantitative trait loci (eQTLs), genetic variants that influence the expression levels of genes. Studying eQTLs is crucial for understanding how genetic variation drives differences in gene activity across tissues and individuals (Nica et al., 2013). GTEx enables the examination of tissue-specific gene regulation, showing how the same genetic variant can affect gene expression differently across organs (Jothi, 2025). This multi-tissue perspective provides deeper biological insight and motivates statistical models capable of handling correlated responses across tissues (Yang, 2021). However, gene expression data are inherently high-dimensional, highly correlated, and often noisy, revealing the necessity for dimension reduction and efficient modeling approaches to capture the potential genetic effects while accurately controlling for variability.

The increase in noisy, high-dimensional data in the biological sciences has highlighted the importance of being able to extract lower-dimensional structure from that data without losing important information. In recent years, there have been many instances of low-dimensional structure providing insight into the underlying biology, both as a tool for understanding and visualization (Simon et al., 2015). Some of the most commonly used dimensionality reduction methods in biological data analysis are Principal Component Analysis (PCA), along with the non-parametric approaches such as t-distributed stochastic neighbor embedding (t-SNE) and uniform manifold approximation and projection (UMAP) (Yan et al., 2021; Yang et al., 2021). Regardless of their underlying properties, whether linear or nonlinear, supervised or unsupervised, these techniques have all proven effective in revealing meaningful structures within high-dimensional data. By applying such methods, researchers can gain deeper insights into complex biological systems more efficiently, without losing critical information. With that in mind, more methods have been developed to deal with specific kinds of datasets while also trying to accomplish the same goal of extracting meaningful statistical information from high-dimensional data (Malepathirana et al., 2022). For example, Vinas et al. (2023) and Yan et al. (2025) employ graph-based factorization and tensor decomposition methods for GTEx data analysis.

Meanwhile, there are a growing number of specialized statistical inference methods for eQTL analyses, for example, quantile regression-based approaches 
(Song et al., 2017; Sun et al., 2019, among others). Linear regression has undergone enhancements specifically tailored for gene expression studies. One such approach is AdaReg (Wang et al., 2020), a data-adaptive robust estimation method developed for linear regression with high-throughput RNA-seq data from the GTEx project. By employing a \(\gamma\)-density-power-weight framework, AdaReg adaptively tunes the \(\gamma\) parameter to balance bias and variance in the presence of gene-specific outliers. This method demonstrated improved estimation performance over fixed-\(\gamma\) and other robust methods, both in simulations and in real GTEx heart tissue data. The results highlight AdaReg’s potential for addressing heterogeneity and outlier contamination in large-scale gene expression analyses. However, AdaReg is primarily developed in a univariate regression setting. Modeling one outcome at a time can ignore correlations among responses and can reduce both power and efficiency in detecting genetic associations (Kim et al., 2023).

To overcome this issue, multivariate regression models can be used to jointly model multiple outcomes, capturing shared genetic effects and improving power in complex genomic settings. However, traditional multivariate approaches often face challenges in adequately modeling cross-tissue dependence when outcomes are correlated. To address this, recent methods such as the multivariate adaptive shrinkage model have been developed, which build on empirical Bayes multivariate regression frameworks to flexibly capture both shared and tissue-specific genetic effects, accommodate missing values, and improve prediction accuracy, particularly in tissues with smaller sample sizes (Morgante et al., 2023). Despite its strength in handling multivariate outcomes, it does not directly address redundant information in modeling high-dimensional GTEx data, where dimension reduction can be beneficial.

To achieve dimension reduction and to further improve estimation efficiency, envelope models have gained significant attention in recent years as a flexible and effective dimension reduction technique, particularly in high-dimensional settings (Cook, 2019). Originally developed for multivariate linear regression (Cook et al., 2010), the method identifies variation that is material to the response while removing immaterial noise, thereby improving estimation and prediction efficiency. Since their introduction, envelope models have been extended to a broad range of statistical frameworks, including well-known methods such as generalized linear models (GLM), partial least squares (PLS), and time series models (Cook and Zhang, 2015; Wang and Ding, 2018; Chakraborty \& Su, 2023, among others), as well as more specialized approaches such as matrix and tensor-variate regressions (Li \& Zhang, 2017; Ding \& Cook, 2018), which allow efficient modeling of complex high-dimensional responses by eliminating redundant variation, and envelope quantile and expectile regressions (Ding et al., 2021; Zhu et al. 2020; Zhao et al., 2022), which enhances estimation robustness by identifying and excluding immaterial information in nonparametric distributional settings.

Recent developments in genetics have applied envelope models to improve estimation efficiency in multivariate regression by accounting for heterogeneity in both regression coefficients and error structures across subgroups. One such study used a Groupwise Envelope Model (Park et al., 2017) on imaging genetics data from the Alzheimer’s Disease Neuroimaging Initiative (ADNI), revealing distinct genetic effects on brain regions between male and female groups. In addition, a Bayesian Simultaneous Partial Envelope Model (Shen et al., 2023) was developed to integrate partial predictor and response enveloping within a unified Bayesian framework, offering flexibility to incorporate prior information and to handle both continuous and discrete predictors. This method also performed well on ADNI data and addressed some limitations of earlier envelope models. These studies illustrate the practical utility and flexibility of envelope models in high-dimensional, heterogeneous genomic contexts, demonstrating their potential to improve estimation and uncover biologically meaningful signals in complex genetic research. Building on this idea, another study proposed a mixture envelope model that integrates envelope modeling with a mixture framework to address latent subgroup structures (Jia, 2018). The method was applied to breast cancer data from TCGA. It evaluated associations between 13 miRNAs and 14 mRNAs to distinguish inflammatory breast cancer (IBC) from non-IBC cases. The model demonstrated improved classification and prediction performance over traditional approaches by effectively capturing subgroup-specific dependency patterns. Other sufficient dimension reduction methods have also been applied to study genomic, multi-omics, and imaging data (Ding \& Cook, 2014, 2015a,b; Jain \& Ding, 2017; Jain et al., 2019; Qian et al., 2019; Ding et al., 2020).

The potential of applying envelope models to GTEx data can help uncover tissue or context-specific genetic effects and biologically meaningful subgroups often overlooked by traditional approaches. As genomic data continue to grow in size and complexity, it becomes challenging to address heterogeneity, high dimensionality, and latent structures in gene expression datasets. To overcome these challenges, we introduce the envelope model as a multivariate dimension reduction strategy for GTEx data analysis. The method can separate variation relevant to gene regulation while filtering out immaterial noise. Correspondingly, it improves statistical power for detecting associations, reduces variance in parameter estimates, and enables more stable inference in high-dimensional settings. Its ability to leverage cross-gene correlations, rather than treating them as nuisance variation, makes the method particularly well suited for GTEx data, as tissue heterogeneity and correlated expression patterns present substantial analytical difficulties. From the study, we demonstrate that envelope models can enhance the detection of significant genes, amplify signals identified by standard linear regression, and uncover additional associations that remain hidden under conventional approaches. Thus, the envelope model offers an efficient and biologically informative modeling framework to investigate GTEx data.

The outline of the rest of the paper is as follows: In Section 2, we introduce the envelope model and its mathematical formulations properties, and estimation algorithm. In Section 3, we employ the envelope model to GTEx data analysis. In particular, in Subsection 3.1, we introduce the background of the GTEx data and the four tissue samples studied in this paper. In Subsection 3.2, we compare the results of the envelope model with the standard multivariate linear model to demonstrate the efficiency gains of the envelope model. In Subsection 3.3, we further evaluate the performance of the envelope model based on similar gene groups and demonstrate its advantages. Finally, Section 4 summarizes the main findings of the study and outlines directions for future research.

\section{ENVELOPE MODEL}
\label{sec:2}
The concept of the envelope model was proposed by Cook et al. (2010), which provides a detailed mathematical and theoretical development of the model. In this section, we provide an introduction to the envelope method.

The envelope model was initially developed based on multivariate linear regression, a fundamental method in multivariate analysis. Consider the standard multivariate linear regression model,
\begin{equation}
  Y = \alpha + \beta X + \varepsilon,
  \label{eq1}
\end{equation}
where \( Y \in \mathbb{R}^r \) is the random response vector,
\( X \in \mathbb{R}^p \) is a non-stochastic vector of predictors,
and the error vector \( \varepsilon \in \mathbb{R}^r \) is normally distributed with mean 0 and unknown covariance matrix \( \Sigma \). Here, \( \alpha \in \mathbb{R}^r \) is the intercept vector and \( \beta \in \mathbb{R}^{r \times p} \) is the coefficient matrix.

The primary goal of the envelope model is to improve the efficiency of estimation in multivariate linear regression by identifying material and immaterial variation in responses in terms of their relevance to predictors and then utilizing only the material information to estimate coefficient parameters. The method is especially beneficial for high-dimensional and correlated responses. Traditional approaches estimate all sources of variation, including noise, which can inflate variance and reduce estimation accuracy. By targeting a minimal subspace, called the envelope, which captures only the material information relevant to the regression, the envelope model achieves substantial gains in efficiency without sacrificing consistency or interpretability. It is particularly valuable in high-dimensional settings such as GTEx data.

Mathematically, the envelope method constructs a link between the mean function and the covariance matrix \( \Sigma \), using the minimal reducing subspace of \( \Sigma \) to cover the regression coefficient space. By employing dimension reduction, the envelope model can reduce the number of parameters in estimation. In addition, the envelope estimator of the unknown coefficient matrix \( \beta \in \mathbb{R}^{r \times p} \) in the multivariate linear regression function has the potential to achieve massive gains in efficiency compared to the standard estimator of \( \beta \), and these gains will be passed on to other tasks, such as prediction and inference. Considering that some characteristics of the response vector may be unaffected by changes in the predictors, enveloping involves making use of the stochastic relationships among the elements of Y and identifying a part of the response that is immaterial to changes in X. More specifically, letting \( (\Phi, \Phi_0) \in \mathbb{R}^{r \times r} \) be an orthogonal matrix. Then Y can be decomposed into two parts:
\begin{equation}
  (i) \; \Phi_0^T Y \mid X \sim \Phi_0^T Y, \; (ii) \; \Phi_0^T Y \perp \Phi^T Y \mid X ,
  \label{eq2}
\end{equation}
where \("\sim" \) means identically distributed, and "\( \perp \)" indicates independence.
Condition (i) implies that the distribution of \( \Phi_0^T Y \) is independent of \( X \), meaning \( \Phi_0^T Y \) carries no direct information about \( \beta \). Condition (ii) indicates that \( \Phi_0^T Y \) is not related to \( X \) through its correlation with \( \Phi^T Y \). Together, these conditions suggest that \( \Phi_0^T Y \) contains no information---either directly or indirectly---about \( \beta \). Accordingly, we refer to \( \Phi_0^T Y \) as the \textit{immaterial} part for the regression, and \( \Phi^T Y \) as the \textit{material} part. These two conditions are algebraically equivalent to:
(a) \( \mathcal{B} \subseteq \text{span}(\Phi) \), where \( \mathcal{B} = \text{span}(\beta) \), the column span of \( \beta \);
(b) \( \Sigma = \Sigma_1 + \Sigma_2 = P_\Phi \Sigma P_\Phi + Q_\Phi \Sigma Q_\Phi \).
Here, \( P_\Phi \) indicates the projection matrix on \( \text{span}(\Phi) \) and \( Q_\Phi = I - P_\Phi =   P_{\Phi_0}\) represents the projection matrix on \( \text{span}(\Phi_0) \), the orthogonal subspace of \( \text{span}(\Phi) \).
When (b) holds, \( \text{span}(\Phi) \) is a reducing subspace of \( \Sigma \) (Conway, 1990; Cook et al. 2010). Thus, (a) and (b) together indicate that \( \text{span}(\Phi) \) is a reducing subspace of \( \Sigma \) that contains \( \mathcal{B} \). To achieve maximum dimension reduction, the envelope of \( \mathcal{B} \), denoted by \( \mathcal{E}(\mathcal{B}) \) or \( \mathcal{E} \), is defined as the smallest reducing subspace of \( \Sigma \) that contains \( \mathcal{B} \). Let \( u = \dim(\mathcal{E}(\mathcal{B})) \) be the envelope dimension (\( u \leq p \)). For notational simplicity, we use \( \Phi \in \mathbb{R}^{r \times u} \) to be a semi-orthogonal basis of \( \mathcal{E}(\mathcal{B}) \). Consequently, \( \mathcal{E}(\mathcal{B}) \) decomposes the total variation into variations related to the material and immaterial parts of \( Y \):
\( \Sigma_1 = \mathrm{Var}(P_\Phi Y \mid X) \) and \( \Sigma_2 = \mathrm{Var}(Q_\Phi Y \mid X) \).

To formulate the envelope model, since \( \mathcal{B} \) is contained in \( \text{span}(\Phi) \), we can write \( \beta = \Phi \eta \), where \( \eta \in \mathbb{R}^{u \times p} \) represents the coordinates of \( \beta \) relative to \( \Phi \). Correspondingly, according to (b), the covariance matrix \( \Sigma \) can be decomposed as \( \Sigma = \Sigma_1 + \Sigma_2  = \Phi \Omega \Phi^T + \Phi_0 \Omega_0 \Phi_0^T \),  where \( \Phi_0 \) is a semi-orthogonal basis of the orthogonal subspace \( \mathcal{E}^\perp \) and is a completion of \( \Phi \), \( \Omega \in \mathbb{R}^{u \times u} (\Omega > 0) \) is the coordinate of \( \Sigma_1 \) relative to \( \Phi \), and \( \Omega_0 \in \mathbb{R}^{(r - u) \times (r - u)}  (\Omega_0 > 0) \) is the coordinate of \( \Sigma_2 \) relative to \( \Phi_0 \). Thus, the envelope model is given by:
\begin{equation}
  Y = \alpha + \Phi \eta X + \varepsilon; \; \; \Sigma = \Phi \Omega \Phi^T + \Phi_0 \Omega_0 \Phi_0^T.
  \label{eq3}
\end{equation}
If \( u < r \), it indicates the presence of immaterial information in the model. By associating the regression coefficients \( \beta \) only with the material part, we eliminate immaterial information and improve the estimation efficiency of the model parameters. In particular, massive efficiency gains can be achieved when \( \Sigma_2 \gg \Sigma_1 \) (Cook et al., 2010). In addition, the envelope model reduces the number of parameters in estimation. In the standard multivariate linear regression  (\ref{eq1}), the total number of parameters includes $r$ parameters for $\alpha$, $rp$ regression coefficients for $\beta$, and $r(r+1)/2$ parameters for the error covariance matrix $\Sigma$, yielding a total of $r + rp + r(r+1)/2$ parameters. In contrast, the envelope model includes $r$ parameters for $\alpha$, $ru - u^2$ parameters for the orthonormal envelope basis $\Phi$, $up$ parameters for $\eta$, $u(u+1)/2$ parameters for the material coordinate matrix $\Omega$, and $(r-u)(r-u+1)/2$ parameters for the immaterial coordinate matrix $\Omega_0$, yielding a total of $r +  up + r(r+1)/2$ parameters. Thus, the envelope model (\ref{eq2}) reduces $(r-u)p$ parameters from (\ref{eq1}). When $u \ll r$, this represents a substantial reduction in model complexity and leads to more efficient estimation. The efficiency gains can be massive when the immaterial variation is large.

Note that the envelope conditions (i) and (ii) (or (a) and (b)) always hold, as a trivial case is the envelope space to be the full space with $u=r$. In this case, \( \mathcal{E}(\mathcal{B}) = \mathbb{R}^r \), implying that no immaterial or redundant information exists, and thus, the envelope model is degraded to and coincides with the standard multivariate linear regression model (\ref{eq1}). For further intuition of the envelope model, see Ding and Cook (2018).

The estimation of the envelope model is based on maximum likelihood estimation (MLE). After a series of derivations, given a basis $\Phi$, the MLEs of $\eta$, $\Omega$, and $\Omega_0$  can all be represented as functions of $\Phi$, and the conditional log likelihood function can be simplified as
\begin{equation}
  L = c - \frac{n}{2}\log (\hat{\Sigma}_Y) - \frac{n}{2}\log |\Phi^T \hat{\Sigma}_{\text{res}} \Phi| - \frac{n}{2} \log |\Phi^T \hat{\Sigma}_Y^{-1} \Phi|,
  \label{eq4}
\end{equation}
where $c= -\frac{nr}{2}(\log(2\pi)+1)$. Then the envelope \( \mathcal{E}(\mathcal{B}) \) can be estimated via Grassmann manifold optimization:
\begin{equation}
  \hat{\mathcal{E}}(\mathcal{B}) = \underset{\text{span}(\Phi) \in G(r, u)}{\arg\min} \left( \log |\Phi^T \hat{\Sigma}_{\text{res}} \Phi| + \log |\Phi^T \hat{\Sigma}_Y^{-1} \Phi| \right)
  \label{eq5}
\end{equation}
where \( G(r, u) \) represents the set of all \( u \)-dimensional subspaces in $\mathbb{R}^{r}$, \( \hat{\Sigma}_Y\) is the sample covariance matrix of \( Y \), and \( \hat{\Sigma}_{\text{res}} \) denotes the sample covariance matrix of the residuals from the standard regression of \( Y \) on \( X \). The Grassmannian optimization can be relatively slow. Therefore, we employ a non-Grassmannian optimization method proposed in Cook et al. (2016).
Without loss of generality, assume that the first \( u \) rows of \( \Phi \), denoted by $\Phi_1$, are non-singular. Then
\begin{equation}
  \Phi = \begin{bmatrix}
    \Phi_1 \\
    \Phi_2
  \end{bmatrix}
  = \begin{bmatrix}
    I_u \\
    A
  \end{bmatrix}
  \Phi_1 = \Phi_A \Phi_1,
  \quad
  \label{eq6}
\end{equation}
where $\quad A = \Phi_2 \Phi_1^{-1}$ is an unconstrained matrix and has $(r-u)u$ free parameters. Then (\ref{eq3}) can be reparameterized into a non-Grassmann optimization as:
\begin{equation}
  \hat{A} = \underset{A \in \mathbb{R}^{(r - u) \times u}}{\arg\min} \;
  -2 \log\left| \Phi_A^T  \Phi_A \right|
  + \log \left| \Phi_A^T \hat{\Sigma}_{\text{res}} \Phi_A \right| \\ + \log \left| \Phi_A^T \hat{\Sigma}_Y^{-1} \Phi_A \right|.
  \label{eq7}
\end{equation}
This can be easily solved with existing optimization methods. Once we obtain the estimator \( \hat A \) or \(\hat{\Phi}_A \), $\hat{\Phi}$ can be estimated by QR decomposition of \(\hat{\Phi}_A \). Then the envelope MLE is \( \hat{\mathcal{E}}(\mathcal{B}) = \text{span}(\hat{\Phi}) \), and the remaining MLEs can be obtained by \( \hat{\eta} = \hat{\Phi}^T \hat{\beta}_{\text{OLS}} \), \( \hat{\beta} = P_{\hat{\mathcal{E}}} \hat{\beta}_{\text{OLS}} =P_{\hat\Phi}\hat{\beta}_{\text{OLS}}  = \hat{\Phi} \hat{\eta}  \), and $\hat{\Sigma} = P_{\hat{\mathcal{E}}}\hat{\Sigma}_{\text{res}} P_{\hat{\mathcal{E}}} + P_{\hat{\mathcal{E}}_0}\hat{\Sigma}_Y P_{\hat{\mathcal{E}}_0} = P_{\hat{\Phi}}\hat{\Sigma}_{\text{res}} P_{\hat{\Phi}} + P_{\hat{\Phi}_0}\hat{\Sigma}_Y P_{\hat{\Phi}_0}= \hat\Phi\hat{\Omega}\hat\Phi + \hat\Phi_0\hat{\Omega}_0\hat\Phi_0 $, where \( \hat{\beta}_{\text{OLS}} \) is the ordinary least squares estimator of \( \beta \) from the standard multivariate linear regression model \eqref{eq1}, $\hat{\Omega} = \hat\Phi^T \hat{\Sigma}_{\text{res}}\hat\Phi$, $\hat{\Omega}_0 = \hat\Phi_0^T \hat{\Sigma}_Y \hat\Phi_0$, and $\hat\Phi_0$ is the orthogonal complement of $\hat\Phi$.

Let
\begin{equation}
  \phi
  =
  \begin{pmatrix}
    \mathrm{vec}(\eta)
    \\
    \mathrm{vec}(\Phi)
    \\
    \mathrm{vech}(\Omega)
    \\
    \mathrm{vech}(\Omega_0)
  \end{pmatrix}
  =
  \begin{pmatrix}
    \phi_1
    \\
    \phi_2
    \\
    \phi_3
    \\
    \phi_4
  \end{pmatrix},
  \label{eq8}
\end{equation}
where the vector operator
$
  \mathrm{vec}
$
yields a vector obtained by stacking the columns of the argument matrix, and $
  \mathrm{vech}
$ applies to a symmetric matrix and stacks only the unique elements of each column of the symmetric matrix. The contraction matrix $
  C_r \in \mathbb{R}^{\,r(r+1)/2 \times r^2}$ and expansion matrix $E_r \in \mathbb{R}^{\,r^2 \times r(r+1)/2}$ satisfy $
  \mathrm{vech}(A)
  =
  C_r \,\mathrm{vec}(A),
$ and $
  \mathrm{vec}(A)
  =
  E_r \,\mathrm{vech}(A),
$ for any symmetric matrix $A$. They are uniquely defined and satisfy $C_rE_r = I_{r(r+1)/2}$. Define

\begin{equation}
  h(\phi)
  =
  \begin{pmatrix}
    \mathrm{vec}(\beta)
    \\
    \mathrm{vech}(\Sigma)
  \end{pmatrix}
  =
  \begin{pmatrix}
    h_1(\phi)
    \\
    h_2(\phi)
  \end{pmatrix}.
  \label{9}
\end{equation}

\begin{proposition}
  Assume $\bar X = 0$. Let $J$ be the Fisher information matrix for\\
  $(\mathrm{vec}(\beta)^T,\mathrm{vech}(\Sigma)^T)^T$
  in the standard model~\eqref{eq1}, and let $V=J^{-1}$.\\ Then
  $\sqrt n (\hat h-h)\overset{d}{\to}N(0,V_0)$,
  where
  $V_0=H(H^TJH)^\dagger H^T$
  and
  $H=\partial h(\phi)/\partial\phi^T$.\\
  Furthermore,
  $V-V_0=V^{1/2}Q_{J^{1/2}H}V^{1/2}\ge0$,
  where
  $Q_{J^{1/2}H}=I-P_{J^{1/2}H}$.
\end{proposition}

The proof of the proposition can be derived similarly from (Cook et al. 2010) and thus is omitted. The proposition theoretically justifies that the envelope estimator is always asymptotically more efficient than or as asymptotically efficient as the standard estimator from the model without enveloping. The efficiency gains can be massive when immaterial variation is large. The dimension $u$ of the envelope model can be chosen based on information criteria such as the Akaike information criterion (AIC) or Bayesian information criterion (BIC), or data-driven cross-validation methods.

\section{GTEx Data Application}\label{sec:data_application}
\subsection{Data Overview and Preprocessing}
In this section, we apply the envelope model to study the GTEx data and compare its performance with the standard multivariate linear regression model. As the GTEx data contain high-dimensional gene expressions, immaterial variation often exists. The envelope model has the advantage of identifying the material and immaterial variations in the data and utilizes only the material information for parameter estimation. By eliminating immaterial variation from model fitting, the envelope method achieves substantial efficiency gains in coefficient estimation and thus identifies more significant genes that cannot be discovered from traditional approaches.

We apply the envelope models to the GTEx data studied by Song et al. (2017), where the authors used data from 4 different tissues in the GTEx data, including 7051 samples of 449 individuals representing 44 tissues (dbGaP accession number phs000424.v6.p1). In our study, we selected the same issues, including muscle-skeletal, whole blood, lung, and thyroid tissues from the GTEx data. Each tissue varies in sample size as well as in gene composition. The selected tissues in this study include muscle-skeletal (\(n = 361\)), whole blood (\(n = 338\)), lung (\(n = 278\)), and thyroid (\(n = 278\)) used for the identification of eQTLs. The gene expressions in the GTEx data have performed several preprocessing steps, including normalization and correction for technical factors. Expression values were transformed to follow approximately normal distributions before statistical modeling. There are also 40 inferred technical covariates, including 3 principal components of SNPs, 35 PEER factors, gender, and genotyping array platform. Since the genotyping array platforms are all the same in the covariate matrices, we only incorporate the remaining 39 inferred covariates (excluding the platform) in our study. We applied the envelope model as a new method to compare its performance with standard multivariate linear regression.

In the envelope model, the predictors are covariate matrices of size \( p \times n \), where \( p \) is the number of predictors, which is 39 in our study, \( n \) is the sample size of each tissue, which is the number of individuals tested for each tissue. The multivariate responses are gene expression matrices of size \(r \times n\), where \(r\) is the number of gene expressions. As discussed in the introduction, the advantage of our method is to provide joint modeling of multivariate genes simultaneously while taking covariances information among the genes into consideration to achieve dimension reduction and efficiency gains. For computational efficiency and ease of implementation, we ranked genes by their marginal expression variance across tissues and retained the top 150 with the highest variance for downstream modeling. This variance-based selection prioritizes genes exhibiting greater biological variability, which are more likely to carry informative signals relevant to genetic regulation. 

\subsection{Envelope Model Implementation and Comparison}
The envelope model is implemented in R using the \texttt{Renvlp} package. The first step is to estimate the dimension of the envelope model using information criteria such as the Akaike information criterion
(AIC) and Bayesian Information Criterion (BIC), or likelihood ratio tests (LRT), or more generally, the cross-validation (CV) method. In our study, we selected the optimal envelope dimension $u$ using BIC. We then fit the envelope model in multivariate linear regression under the optimal envelope dimension \( u \). The results with other dimension selection methods are generally consistent. The default output contains the following components:
\begin{itemize}
  \item \( \hat{\beta} \): The envelope estimator  $\hat{\beta} = \hat{\Phi} \hat{\eta}$.
  \item \( \hat{\Sigma} \): The envelope estimator $\hat{\Sigma} = \hat\Phi\hat{\Omega}\hat\Phi + \hat\Phi_0\hat{\Omega}_0\hat\Phi_0$.
  \item \( \hat\Phi \): The envelope estimator of an orthonormal basis of the envelope subspace $\mathcal{E}$.
  \item \(\hat \Phi_0 \): The envelope estimator of an orthonormal basis of the complement of the envelope subspace $\mathcal{E}_0$.
  \item \( \hat\eta \): The envelope estimator of the coordinates of \(\beta\) with respect to \(\Phi\).
  \item \( \hat\Omega \): The envelope estimator of the coordinates of \(\Sigma\) with respect to \(\Phi\).
  \item \( \hat\Omega_0 \): The envelope estimator of the coordinates of \(\Sigma\) with respect to \(\Phi_0\).
  \item \( \hat{\alpha} \): The estimated intercept.
  \item \texttt{asySE}: The asymptotic standard errors for elements in \(\beta\) and \(\Sigma\) under the envelope model.
\end{itemize}

After building the envelope model, we need to obtain evidence to determine whether the envelope model is more effective in analyzing the GTEx data. To quantify this efficiency gain, we computed the asymptotic standard error ratio for each estimated regression coefficient in $\beta$, defined as the asymptotic standard error under the standard multivariate linear regression model \eqref{eq1} divided by the corresponding asymptotic standard error under the envelope model \eqref{eq2} and denoted as $SE_{\hat\beta_{OLS}} / SE_{\hat\beta}$, or $SE_{LM} / SE_{ENV}$. Larger ratios indicate greater efficiency gains achieved by the envelope model. The results show that the majority of these ratios are greater than one across tissues and covariates, with median values exceeding one in all cases, implying that the envelope model generally provides more stable and precise coefficient estimates and improves the capacity to detect gene-expression biomarkers.

In particular, the efficiency ratios shown in Figure 1 demonstrate that the envelope model generally produces smaller standard errors than the standard linear model across most tissues and covariates. For the gender covariate, the ratios are all greater than one, particularly in muscle and thyroid tissues, indicating substantial efficiency gains. Lung and whole blood tissues exhibit more moderate improvements, though the median ratios still exceed one. Similar patterns are observed for the principal component covariates. For both PC1 and PC2, the envelope model again shows improved efficiency across all tissues, with especially notable gains in muscle, thyroid, and whole blood. In contrast, lung tissue consistently shows smaller values in all covariates but still positive improvements. Overall, the detection of additional significant genes under the envelope model highlights the practical advantage of variance reduction in improving statistical power for gene expression studies because it provides more precise coefficient estimates for gene expression analysis across multiple covariates and tissues.

\begin{figure}[h]
  \centering
  \includegraphics[width=0.60\textwidth]{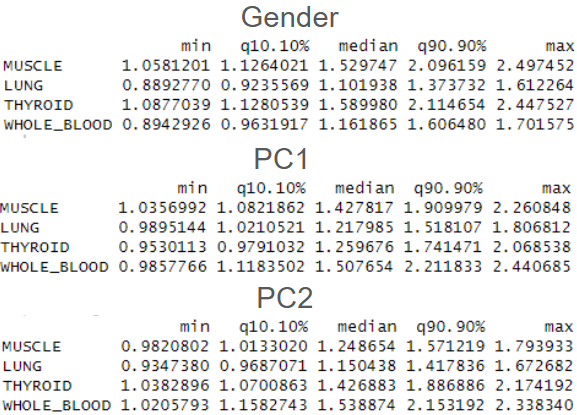}
  \caption{Summary statistics of the efficiency ratios $SE_{\hat\beta_{OLS}} / SE_{\hat\beta}$  across genes for each tissue and covariate (Gender, PC1, and PC2). Ratios greater than one indicate that the envelope model produces smaller standard errors than the standard linear model, reflecting improved estimation efficiency. The table reports the minimum, 10th percentile, median, 90th percentile, and maximum values of the ratios for each tissue.}
  \label{fig:four_tissues}
\end{figure}

In addition, consistent patterns were observed in the hypothesis testing and the corresponding $p$-values of the estimators obtained from both the envelope model and the standard linear model. Since gene expression studies involve testing a large number of genes simultaneously, multiple hypothesis testing can substantially increase the probability of false positives. To address this issue and control the false discovery rate (FDR), we applied the FDR adjustment implemented in R to compute adjusted p-values. Specifically, the procedure was used to control the expected proportion of falsely rejected hypotheses among all significant findings. This adjustment provides a more reliable assessment of statistical significance when evaluating gene–covariate associations across thousands of genes.

Figure 2 shows the statistical significance of the gender effect on gene expressions across four tissues. Each gray bar corresponds to a gene, and the height of the bar represents the value of $-\log_{10}(p)$ for the estimated gender coefficient, where $p$ represents the $p$-value of the coefficient. Smaller $p$-values indicate stronger evidence that gene expression is associated with gender. The red horizontal line denotes the significant threshold corresponding to $p=0.05$ after adjustment, such that genes with bars extending above this line are considered statistically significant. The left panel displays the results from the envelope model (ENV), and the right panel shows results from the standard linear model (LM).

The genes that are highlighted in blue in Figure 2 are identified as significant by the envelope model, but not by the standard linear model, indicating that additional discoveries are enabled by the envelope approach. To determine whether these findings correspond to biologically meaningful signals rather than random statistical fluctuations, we reviewed the existing literature for several representative genes. As discussed below, many of the genes that are uniquely captured by the envelope model have been proven to associate with sex-dependent expression patterns or participate in biological pathways known to differ between males and females. These findings provide external biological support for the additional signals uncovered by the envelope model.

In particular, in the muscle tissue, LSMEM2 has been recently characterized as a membrane protein expressed in skeletal muscle. Several studies demonstrate that overexpression of LSMEM2 leads to abnormal muscle phenotypes, and these functional abnormalities were observed in male mice, suggesting a potential link to sex-dependent physiological processes (Elrefaei et al., 2025). In addition, CISH, a key regulator in the JAK–STAT signaling pathway, is responsive to growth hormone (GH) stimulation in skeletal muscle. It is known to differ between males and females due to hormonal regulation; the induction of CISH in muscle highlights its potential role in sex-dependent metabolic processes (Clasen et al., 2013). Similarly, gene FGF18, involved in skeletal muscle physiology, demonstrates sex-dependent expression responses under metabolic stress conditions such as fasting (Jia et al., 2020).

A similar pattern is noted in lung tissue, where KIF26B, HHLA2, and PCP4L1 are found to be specifically detected by the envelope model. Both KIF26B and HHLA2 are associated with immune-related pathways (Cheng et al., 2017; Sun et al., 2022), with HHLA2 playing a crucial role in immune evasion mechanisms involving T-cell metabolism. Since lung immune responses, particularly those involving T-cell activity, are known to exhibit sex-dependent differences, the relevance of these genes is further supported (Brown et al., 2020; Puttur et al., 2026). In addition, PCP4L1 is implicated in transcriptional responses to environmental stressors, where transcriptomic analyses of lung tissue have revealed distinct sex-specific gene expression patterns across multiple cellular compartments (Choudhary et al., 2020).

TFEC, TNFRSF18, and CHAC1 were significant in the thyroid tissue under the gender category by the envelope model but not by the linear model. This makes biological sense as recent studies have shown that thyroid gene regulatory networks in thyroid tissue vary in a sex-dependent manner, particularly in immune and metabolic pathways (Jia et al., 2025; Saha et al., 2025; Liu et al., 2025). These results may indicate that the envelope model could be more sensitive in detecting biologically significant signals related to sex.

In whole blood tissue, genes such as RGS9, MYL3, and RHAG are found to be significant only when the envelope model was applied. Sex differences in the blood immune systems are well documented. There are differences in terms of variations in gene expression profiles, immune cell composition, and inflammation between males and females (Bongen et al., 2019; Paltiel et al., 2025; Sharma et al., 2025). Considering this background, the presence of such genes in gender-associated analysis may suggest the presence of important signals by the envelope model.

Overall, these findings suggest that the envelope model may offer increased capacity/sensitivity in detecting biologically relevant gender-associated biomarkers that may be missed by conventional approaches. Rather than finding random additional genes, the envelope model consistently points to genes associated with pathways known to have gender-dependent regulation.

\begin{figure}[h]
  \centering
  \includegraphics[width=0.60\textwidth, height=.3\textheight]{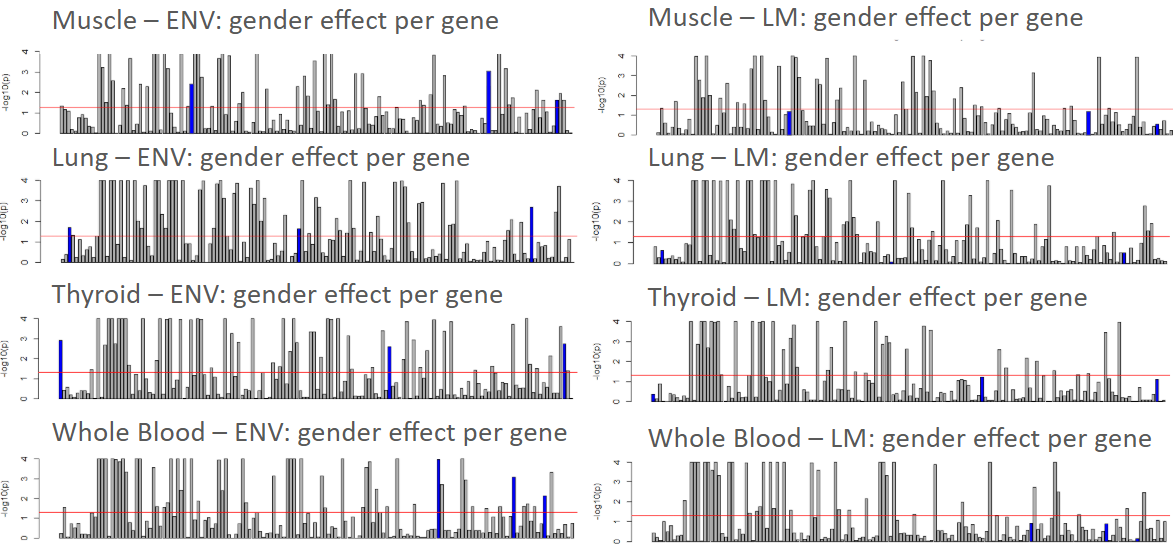}
  \caption{Bar plots comparing the statistical significance of gender effects on gene expressions across four different tissues (Muscle, Lung, Thyroid, and Whole Blood) between the envelope model and standard model. The left panels correspond to the envelope model (ENV), while the right panels correspond to the standard linear model (LM).}
  \label{fig:four_tissues}
\end{figure}

The results based on the gender effects provide external validation for the proposed approach, as many of the genes that were specifically identified by the envelope approach have been previously identified by relevant biological studies. Although the principal components (PC1 and PC2) do not map to immediately interpretable variables, they do represent the major modes of variation present in the data. In Figure 3 and Figure 4, the envelope approach consistently identifies a larger number of significant genes than the standard linear model. These results provide converging evidence that the envelope approach provides increased capacity for the detection of meaningful signals, both for interpretable variables and for latent variables. 

\begin{figure}[h]
  \centering
  \includegraphics[width=0.60\textwidth,height=.3\textheight]{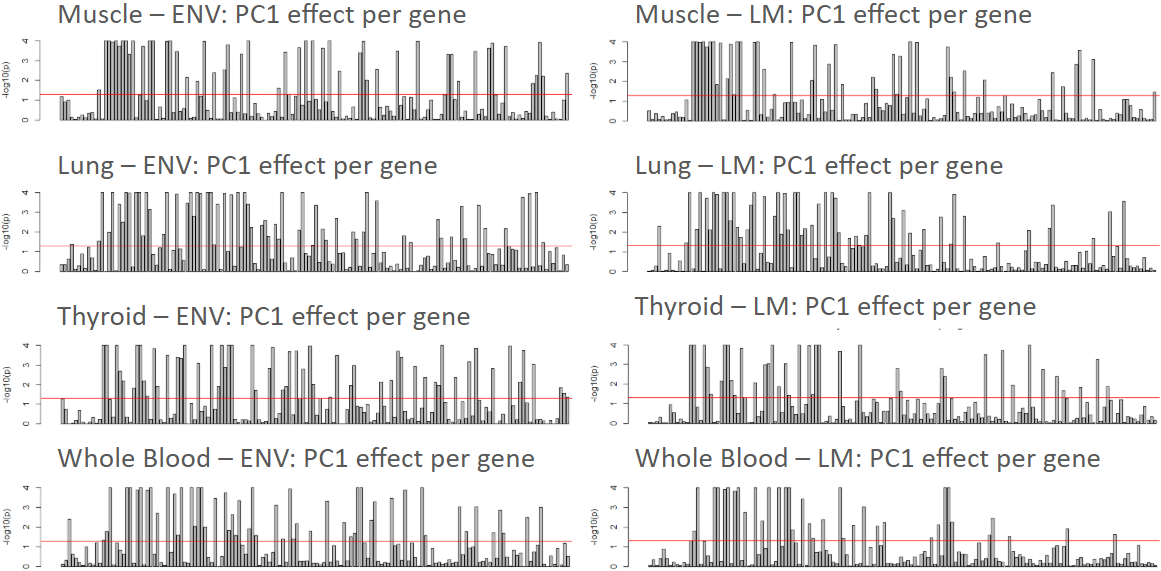}
  \caption{Bar plots comparing the statistical significance of associations between gene expression and the first principal component (PC1) across four tissues (Muscle, Lung, Thyroid, and Whole Blood) between the envelope model and standard model. The left panels correspond to the envelope model (ENV), while the right panels correspond to the standard linear model (LM).}
  \label{fig:four_tissues}
\end{figure}

\begin{figure}[h]
  \centering
  \includegraphics[width=0.60\textwidth,height=.3\textheight]{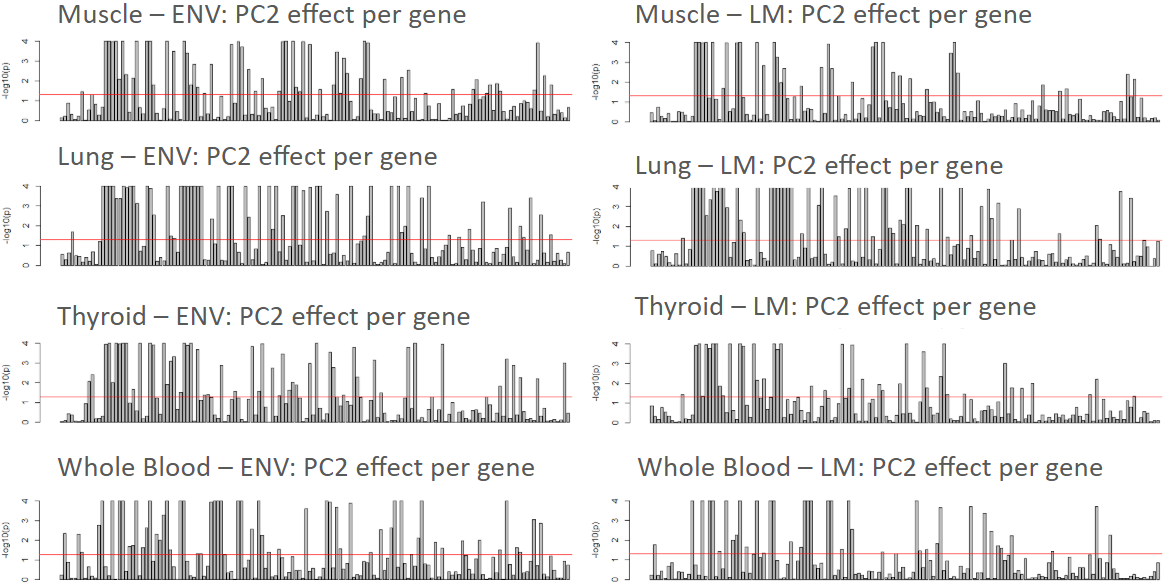}
  \caption{Bar plots illustrating the significance ($-\log_{10}(p)$) of associations between gene expression and the second principal component (PC2). The Envelope Model (ENV, left) and Linear Model (LM, right) are shown for each tissue.}
  \label{fig:four_tissues}
\end{figure}

In summary, Figures 2–4 consistently demonstrate that the envelope model improves detection power across all tissues while maintaining the same level of multiple-testing control. Compared to the standard linear model, the envelope model approach uncovers substantially more genes associated with the biological covariate gender and the dominant sources of transcriptomic variability captured by PC1 and PC2. Table 1 provides a quantitative summary of the number of significant genes identified by the envelope model and the standard model across the four tissues for Gender, PC1, and PC2. For every tissue examined, the envelope model detects a larger number of significant genes than the linear model. Meanwhile, a substantial proportion of the genes identified by the linear model are also detected by the envelope model, as reflected by the large overlap counts in the ``Common" column. This indicates that the envelope model retains most of the discoveries made by the standard approach while simultaneously identifying additional significant genes that are not detected by the linear model. The consistent increase in the number of significant findings across tissues suggests that the envelope model achieves greater statistical efficiency by reducing immaterial variation through dimension reduction, therefore improving statistical power in hypothesis testing and detection capacity.

\begin{table}
  \caption{Summary of the number of the significant genes ($p < 0.05$) for associations with Gender, PC1, and PC2 across four tissues, identified by the envelope and standard models. The ``Common'' column indicates genes identified as significant by both models.}
  \label{tab:four_tissues}

  \begin{tabular*}{\textwidth}{@{\extracolsep{\fill}}lccccccccc}
    \hline\noalign{\smallskip}
    & \multicolumn{3}{c}{Gender}
    & \multicolumn{3}{c}{PC1}
    & \multicolumn{3}{c}{PC2} \\
    \cline{2-10}
    Tissue & ENV & LM & Common & ENV & LM & Common & ENV & LM & Common \\
    \noalign{\smallskip}\svhline\noalign{\smallskip}

    Muscle  & 56 & 45 & 38 & 51 & 42 & 37 & 61 & 36 & 35 \\
    Lung    & 63 & 49 & 44 & 64 & 55 & 51 & 65 & 59 & 52 \\
    Thyroid & 61 & 41 & 36 & 59 & 40 & 37 & 53 & 42 & 36 \\
    Blood   & 55 & 41 & 39 & 55 & 37 & 32 & 57 & 40 & 35 \\

    \noalign{\smallskip}\hline\noalign{\smallskip}
  \end{tabular*}

\end{table}

\FloatBarrier
\subsection{Additional Analysis Based on Similar Gene Groups}
To further investigate the structured variation in gene expression data, we conducted additional analysis based on similar gene groups. In particular, genes that share common prefixes were generally grouped into clusters because naming patterns often reflect shared genomic annotations or related transcript families. This grouping provides a practical approach to partitioning genes into subsets with potentially similar biological characteristics.

Using the muscle-skeletal tissue dataset, we selected two representative gene groups of different sizes for analysis. The corresponding dataset contains 491 samples, and the covariate matrix includes 64 inferred covariates, comprising 3 principal components derived from SNP data, 60 PEER factors, and gender. The first group consists of 261 genes with names beginning with the prefix “CTD”, while the second group includes 73 genes with names beginning with “AP0”. The envelope modeling procedure presented in Section B was applied to each cluster separately. The dimension of the envelope subspace was selected via BIC with an estimated dimension of 45 for Group I and 14 for Group II.

Similarly, to assess the potential efficiency gains of the envelope model, we computed the asymptotic standard error ratio for each estimated regression coefficient in $\beta$, defined as the asymptotic standard error under the standard multivariate linear regression model divided by the corresponding asymptotic standard error under the envelope model. This yielded a $r \times p$ ratio matrix, where each entry represents the ratio of the asymptotic standard error from the standard multivariate linear regression model to the corresponding asymptotic standard error from the envelope model. Larger ratios indicate greater efficiency gains achieved by the envelope model. For illustration, we randomly selected three rows and columns (denoted as A, B, and C) from the ratio matrix. Each column of the ratio matrix represents the standard error ratios (efficiency gains) for the estimated effects of a given covariate on all response genes, while each row contains the standard error ratios (efficiency gains) for the estimated effects of all covariates on a particular gene. 

As summarized in Table 2, all the ratios have values above one, ranging between 1.00 and 4.91 across both gene groups. Since values greater than one indicate that the envelope model produces lower asymptotic standard errors than the standard multivariate linear regression model, these results show consistent efficiency gains from the envelope approach. The mean ratios are generally between 2 and 3, implying that the standard errors obtained from the linear model are typically two to three times larger than those obtained from the envelope model. In some cases, the ratio approaches 5, indicating massive efficiency gains in estimation efficiency. These improvements arise from the ability of the envelope model to remove immaterial variation that does not contribute to the regression signal while retaining only the variation relevant to gene-covariate associations in estimation. The reduction in estimation variability helps explain the efficiency gains of the envelope model observed in the analyses.

\begin{table}
  \caption{Average, maximum, and minimum values of the standard error ratio $SE_{LM}/SE_{ENV}$ for selected rows and columns of the coefficient matrix in the two gene clusters.}
  \label{tab:se_ratio}

  \begin{tabular*}{\textwidth}{@{\extracolsep{\fill}}cccccccc}
    \hline\noalign{\smallskip}
    Group & Level & Row A & Row B & Row C & Col A & Col B & Col C \\
    \noalign{\smallskip}\svhline\noalign{\smallskip}

    I
    & Mean & 1.9869 & 1.9798 & 2.0166 & 2.4426 & 2.3944 & 2.3462 \\
    & Max  & 3.8642 & 3.6484 & 3.9664 & 3.9664 & 3.8183 & 3.6492 \\
    & Min  & 1.0051 & 1.0179 & 1.0191 & 1.3391 & 1.3158 & 1.3121 \\

    \noalign{\smallskip}\hline\noalign{\smallskip}

    II
    & Mean & 2.9070 & 3.2206 & 3.0882 & 2.3940 & 2.5833 & 2.5466 \\
    & Max  & 4.3976 & 4.3420 & 4.4857 & 3.9972 & 4.9156 & 4.7101 \\
    & Min  & 1.0952 & 1.3044 & 1.1478 & 1.1378 & 1.1595 & 1.1553 \\

    \noalign{\smallskip}\hline\noalign{\smallskip}
  \end{tabular*}

\end{table}

In addition, we computed FDR-adjusted $p$-values on a $-\log_{10}(p\text)$ scale for both gene groups under models with and without enveloping. The findings are illustrated in Figures 5 and 6. The red horizontal line indicates the statistical significance threshold corresponding to $p = 0.05$. Similar to the results presented in Section B, for both groups, there are a higher number of statistically significant coefficients detected by the envelope approach than by the linear regression method.

\begin{figure}[H]
  \centering
  \includegraphics[width=0.80\textwidth,height=.3\textheight]{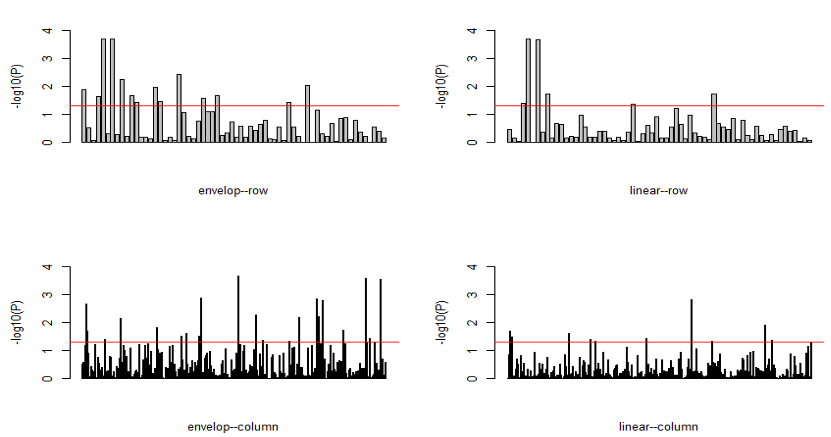}
  \caption{The FDR-adjusted $p$-values (on a -\( \log_{10} \) scale) for the estimators of the covariate effects on Group I gene expressions from the envelope model and the model without enveloping. The estimators are from a randomly selected column and row in the coefficient matrix $\beta$ with and without enveloping; the results from other columns and rows are similar.}
  \label{fig:cluster1}
\end{figure}

\begin{figure}[h]
  \centering
  \includegraphics[width=0.80\textwidth,height=.3\textheight]{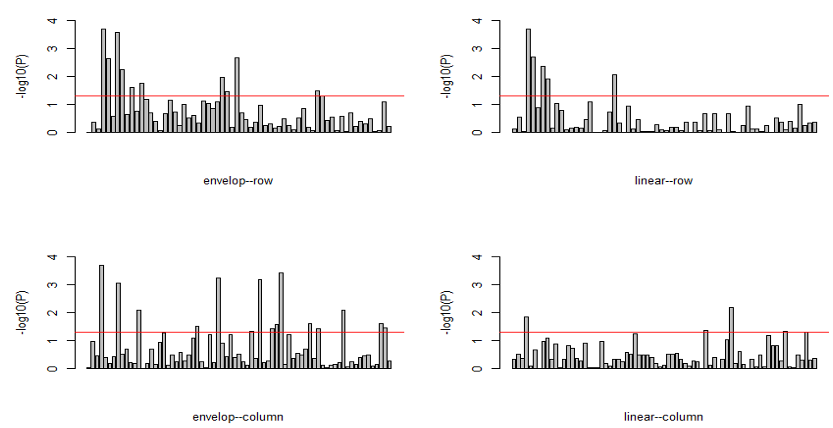}
  \caption{The FDR-adjusted $p$-values (on a -\( \log_{10} \) scale) for the estimators of the covariate effects on Group II gene expressions from the envelope model and the model without enveloping. The estimators are from a randomly selected column and row in the coefficient matrix $\beta$ with and without enveloping; the results from other columns and rows are similar.}
  \label{fig:cluster2}
\end{figure}

Overall, the group-based analysis further supports the advantages of the envelope model by reducing immaterial variation within gene subsets. Therefore, the findings from both the data analyses in Section B and C provide coherent evidence that the envelope model achieves more efficient performance relative to the standard approach. This demonstrates the robustness of the envelope model across different underlying data structures, including both heterogeneous gene expression data and more homogeneous groups of genes.

\FloatBarrier
\section*{IV. DISCUSSION AND CONCLUSION}
In this study, we investigated the advantages of the envelope model for gene expression analysis using GTEx data. Compared with standard multivariate linear regression, the envelope model consistently identified a larger number of significant genes across all four tissues (muscle-skeletal, lung, thyroid, and whole blood), while retaining the genes detected by the standard model. By effectively separating material and immaterial variation, the envelope model substantially reduced the standard errors of the estimated effects and obtained notable efficiency gains and stronger statistical signals. These reductions in estimation variability translated directly into increased power for hypothesis testing and enabled the detection of additional biologically meaningful associations without sacrificing the discoveries obtained from the standard approach. Further analysis based on similar gene groups reinforced these findings. By applying the envelope model within each gene group, we observed substantial efficiency gains via enveloping. 
These results highlight the capacity of the envelope model to amplify biologically relevant signals that traditional approaches may overlook, and suggest that the envelope method improves statistical efficiency and power in estimation and inference.

In conclusion, our findings in the GTEx analysis underscores the advantages of envelope models for genomic studies by providing more stable inference and improved detection of tissue- and context-specific effects. The envelope method yields more efficient, interpretable, and biologically meaningful results, and demonstrates clear methodological and practical benefits. Its flexibility in handling tissue-specific and high-dimensional genomic data makes it a valuable complement to existing GWAS, eQTL, and TWAS approaches for studying genetic regulation and complex traits.

Future research directions include extending the envelope method to directly model highly skewed and zero-inflated count data, which frequently arise in genomic studies, and to address increasingly complex data structures, including single-cell and spatial transcriptomics, multi-omics integration, and longitudinal molecular profiling studies. In addition, scalable computational algorithms for envelope and regularized envelope approaches are needed to handle the rapidly growing dimensionality of modern genomic datasets.


\end{document}